# Direction-sensitive graphene flow sensor

P. Kaźmierczak [1], J. Binder [1], K. Boryczko [1], T. Ciuk [2], W. Strupiński [3], R. Stępniewski [1] and A. Wysmołek [1]

[1] Faculty of Physics, University of Warsaw, Pasteura 5, 02-093 Warsaw, Poland

[2] Łukasiewicz Research Network – Institute of Microelectronics and Photonics, Aleja Lotnikow 32/46, 02-668 Warsaw, Poland

[3] Physics Faculty, Warsaw University of Technology, Pl. Politechniki 1, 00-661 Warsaw, Poland

Abstract:

Graphene flow sensors hold great prospects for applications, but also encounter many difficulties, such as unwanted electrochemical phenomena, low measurable signal and limited dependence on the flow direction. This study proposes a novel approach allowing for the detection of a flow direction-dependent electric signal in aqueous solutions of salts, acids and bases. The key element in the proposed solution is the use of a reference electrode which allows external gating of the graphene structure. Using external gating enables to enhance substantially the amplitude of the flow-generated signal. Simultaneous measurement of the reference electrode current allows us to recover a flow-direction-sensitive component of the flow-induced voltage in graphene. The obtained results are discussed in terms of the Coulomb interaction and other phenomena which can be present at the interface of graphene with the aqueous solution.

The remarkable properties of two-dimensional (2D) materials, such as high chemical and physical resistance to harsh environment as well as excellent electrical properties, make them a highly interesting subject for many researchers. For more than two decades, 2D materials are the subject of studies that aim to develop novel applications [1-7]. As an archetype for the 2D materials family, graphene was considered a promising candidate for many applications, including liquid flow sensors. Since the work of Král and Saphiro in 2001 [8], which shows the generation of electricity in a system with carbon nanotubes immersed in a flowing liquid, there have been many papers aimed at explaining the effects occurring in various carbon materials (later also 2D materials, including graphene), exposed to a liquid. Various theories have been proposed to explain the observed impact of liquid flow on carbon-based nanostructures, for example phonon coupling, Coulomb interactions [9-20] streaming potential [21-23], capacitive discharge, waving potential [24-27] and other [28-33]. In spite of the fact that the current understanding of the electric signal generation by flowing liquid is still very limited, we propose an approach which allows substantial enhancement of the flow-generate signal as well as a direction-sensitive detection of the flow of aqueous solutions.

One of the main problems of the systems aimed to measure flow-induced signals is the presence of many electrochemical phenomena which can change the signal diametrically, being up to 2-3 orders of magnitude larger than the flow signal. In our work, we propose a measurement method that allows us to separate different electrical currents, especially those which depend and those that do not depend on the flow direction - thus eliminating the signal of electrochemical origin. This is very promising for the application of graphene flow sensors on the macro and microscale.

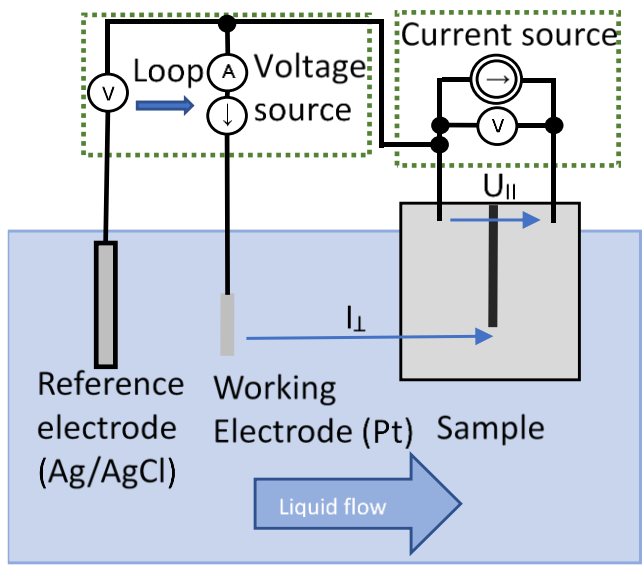

Fig. 1. Schematic drawing of the measurement setup.

Epitaxial graphene was grown by Chemical Vapor Deposition (CVD) or the sublimation method on 4H-SiC insulating substrates [34]. Both Si- and C-faces were used. The quality of graphene was studied by Raman spectroscopy. The FWHM of the 2D band was in the range of 50-65 cm$^{-1}$, fitted by a single Lorentzian function, which corresponds to mono- and bi- layer of epitaxial graphene [35]. The ratio of the amplitude of the D band to the G band, which can be used to estimate the number of defects in the sample, lies in the range of $I_D/I_G$ ~0.2 – 0.6, which indicates a high quality of the investigated material [36]. We examined several few layer graphene samples with different electrical properties such as sheet resistance (2 – 12k Ω/sq.), carrier concentration: n-type (1.5 x 10$^{12}$ – 7.0 x 10$^{12}$ cm$^{-2}$) and p-type (~1.5 x 10$^{13}$ cm$^{-2}$) and carrier mobility in the range 250 – 3200 cm$^2$/Vs. Most of the experiments were carried out on a sample that was produced by the CVD method and had the initial parameters: mobility μ~600 cm$^2$/Vs and electron concentration n = 3.3x10$^{12}$ cm$^{-2}$.

The passivated contact pads on the top of the sample are connected to the active area on the bottom part via graphene conductive channels. These channels are made by etching a graphene strip which electrically separates the two leads from the contact pads. This way, the active part of the sample can be immersed into the liquid and at the same time one can avoid an electrical connection between the metallic contacts, which remain above the liquid level (sample in Fig. 1).

The measurement system (see Fig. 1.) is based on a glass vessel placed on a turntable and a sample immersed in a solution. The movement of the rotating table causes a movement of the liquid with regard to the stationary sample. Due to the much larger diameter of the vessel relative to the sample, it can be assumed that the movement of the liquid near the sample was a laminar flow. For electrical measurements the system is equipped with a source measure unit (Agilent B2901A) and two additional electrodes: a reference and a working electrode. As the reference, an Ag/AgCl electrode was used to ensure a constant reference potential in each measurement. During the experiment we applied a voltage $V_G$ between the Pt electrode and the sample and measured a current $I_\perp$, which flows perpendicularly to the liquid flow. The voltage $U_{||}$, was measured on the contacts above the liquid level with constant forced current typically of the order of tenth of μA. It was found that keeping a constant forced current trough the sample, measured results are stabilized probably due to elimination of the parasitic, unstable currents caused by non intentional contact asymmetry, temperature gradients etc. The change in the direction of the liquid flow was obtained by changing the direction of rotation of the turn table. The reference measurement of the rotation speed is carried out by a calibrated tachometer. The flow velocity is calculated as: $v = \omega \cdot r$, where $r$ is distance of the sample from the axis of rotation, ω is rotation speed.

It has been shown that unpassivated metal contacts can lead to unwanted electrochemical effects and should therefore be avoided [17]. We circument this problem by using a special sample design shown in Fig. 1.

A typical $U_{||}$ measurement is presented in Fig. 2. The observed signal variations are caused by changes in the liquid flow velocity and direction. As one can see, there is a clear difference in the measured voltage $U_{||}$ between flow "on" and "off" states. At the interface between the electrolyte and the semiconductor the electric double layer is formed. Changes in the distribution of charges in the double layer caused by switching on the liquid flow cause a different level of the $U_{||}$ signal ($U_R$ and $U_L$ in Fig. 2). The change in the level of the $U_{||}$ signal for changing the direction of the liquid flow is also clearly visible ($U_{L/R}$ in Fig. 2). Similar to the case discussed in the literature [37] we observe a non-zero voltage for the lack of liquid flow. This is the result of inhomogeneities occurring both on the graphene sample and the measuring contacts as well as the applied voltage $V_G$ and current $I_\perp$.

As shown in Fig. 2, some signal components depend on the velocity direction of the liquid and some of them occur even when the flow is switched off. We assume that the signal itself can be divided into three components: (1) Background signal - which is always present, even when the liquid flow is turned off. (2) Flow-related signal - occurs upon liquid flow. (3)

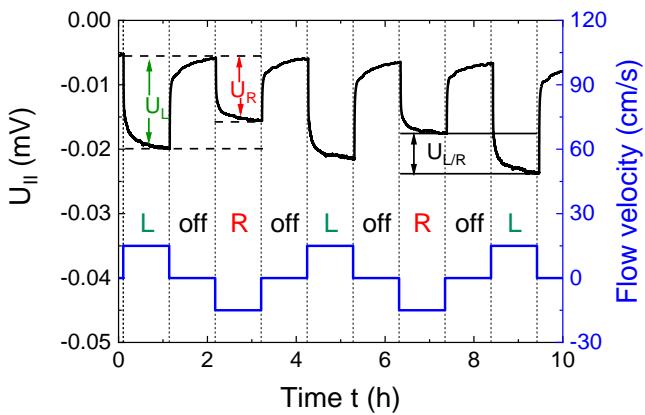

Fig. 2 Voltage $U_{II}$ measured as a function of the velocity and direction of the liquid flow ($V_G$= -0.2 V). The flow velocity of the liquid is indicated on the right blue scale. The area marked as "off" means no flow. The areas with the green "L" and the red "R" differ in the direction of the liquid flow. $U_R$ and $U_L$ correspond to the difference in signal level between the "on" and "off" regions for right and left flow direction, respectively. $U_{L/R}$ is the difference between $U_R$ and $U_L$.

Direction-sensitive signal - occurs when the liquid flow is turned on and is responsible for the difference in the signal between two liquid flow directions.

A set of experimental results obtained for two flow directions (L- left, R-right) for different applied voltages $V_G$ between the sample and the working electrode are presented in Fig. 3. For different values of $V_G$, which modify the charge distribution in the double layer, we observe a different response of the system. For each $V_G$ value, we observe a different value for the "off" signal level, a different value of $U_R$ and $U_L$ and different times of signal stabilization. The highest value of $U_R/U_L$ is observed for a value of $V_G$ = -0.4 V. However, for this value, the time needed for the signal to stabilize is also the longest. For $V_G$ = 0 V, we have the opposite situation. In order to get a stable and strong signal, an appropriate $V_G$ must be selected. Another important observation is the change of the generated signal for the values of $V_G$ = 0.1 V, and 0 V, as shown in Fig. 4. It is clearly observable that the signal for $V_G$ = 0 V is reversed with regard to $V_G$ = 0.1 V. This effect is related to the change of the sign of charges in the double layer. For this value, the $I_L$ current also changes its sign.

The $V_G$ dependence shown in Fig. 3 illustrates that for certain voltages ($V_G$), very weak or for some samples even no flow effects can be observed. Almost in every recorded measurement, external voltages had to be used to search for an appropriate charge distribution in the double layer in order to obtain a useable flow dependent signal. Without additional voltage $V_G$, the vast majority of measurements showed only noise, in agreement with Yin et al. [17]. The authors of work [37] observed similar effects on devices based on carbon nanotubes. However, these studies [37] were intended to explain the nature of the signal generation phenomenon, and the fundamental importance of an additional voltage control ($V_G$) to obtain measurable effects on a wide range of samples has not been recognized. We think that changing the charge distribution, for example, by using an external voltage, is an essential element to obtain measurable flow effects in such systems.

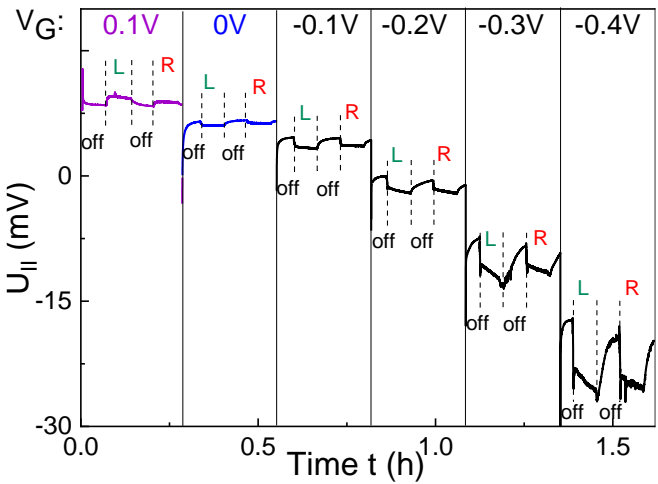

Fig. 3 Dependence of the generated signal $U_{II}$ on a change of the applied voltage $V_G$.

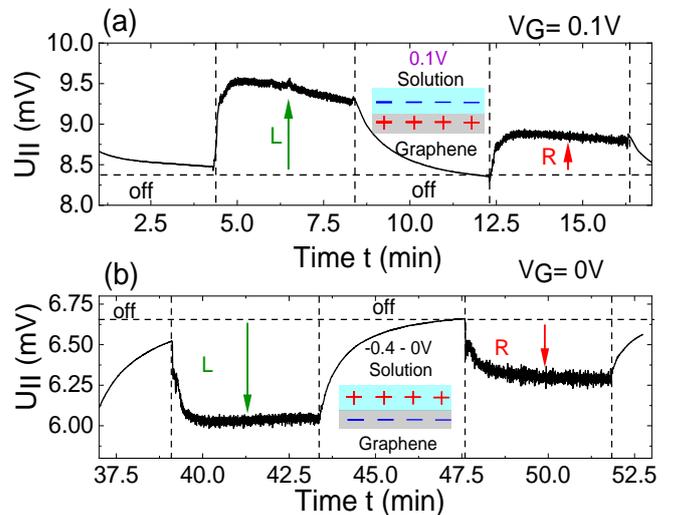

Fig. 4 Graphs show enlarged parts of Fig. 3 (a) $V_G$ equal to 0 V and (b) 0.1 V. On the graphs, there are schematic drawings illustrating the charge distribution within graphene (grey part) and on its surface exposed to the solution, for respective $V_G$.

The measured signals $I_\perp$, like the $U_\parallel$ signal from the sample, depends on the parameters of the double layer. This effect can be illustrated by comparing both recorded signals ($I_\perp$, $U_\parallel$) in one figure (Fig. 5 upper part). The two charts are almost alike. Both signals respond to most of the changes in the same way. Both signals appear to be proportional to each other, with only incremental differences that depend on different minor components of the signal. This gives us the opportunity to compare and subtract these two signals and extract direction sensitive part of the signal:

$$\Delta U = U_\parallel - I_\perp \alpha - \beta \qquad (1)$$

Both measurements $I_\perp$ and $U_\parallel$ differ in the measured unit Ampere and Volt, respectively. To be able to subtract them, these units should be unified. The value of scaling factor α is selected so that the two signals are as close as possible in value. Due to the applied external voltage source and built-in potential differences, sometimes one of the signals needs to be additionally shifted by a certain constant β, so that the value of ΔU is equal to 0 V for lack of flow. The result of the subtraction is shown in Fig. 5 – lower part. As a result, we obtain a signal ΔU which is close to zero for the situation without any flow and non-zero in the case of a liquid flow. Most importantly, sign of the obtained signal shows a dependency on the direction of flow. Positive values indicate flow to the right, and negative values indicate flow to the left. We believe that this observation of a direction dependent signal confirms the existence of a drag effect.

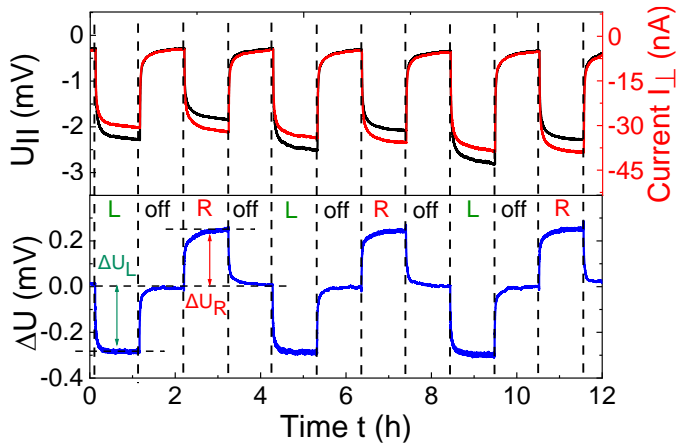

Fig. 5 The upper part shows the detected signal on the graphene sample $U_\parallel$ (left axis) and the perpendicular current through the solution $I_\perp$ (right axis). The lower part shows the result after applying a subtraction procedure of $V_\parallel$ and $I_\perp$.

The changes caused by switching the flow on and off and the change of the flow velocity for $U_\parallel$ are at least an order of magnitude lager than the changes for the signal obtained after subtracting the signals $U_\parallel$ and $I_\perp$ from each other - ΔU. When considering the use of such a system for a flow sensor that only detects flow velocity or only its appearance, one can rely on the $U_\parallel$ value. However, if we need to know the flow direction in addition, we should also use the calculated value of ΔU.

The time scale of the experiment in Fig. 5 is expressed in hours. One measurement cycle consisting in switching the flow on and off, lasted 2 hours. During this time, the signal still stabilizes, not reaching a constant value. However, the largest changes in the generated signal happen on a faster time scale. A 50% change in measured voltage value is achieved within 30 seconds. This is due to the use of the system with a rotary table, which causes the liquid flow velocity to stabilize after several seconds, we can correctly interpret the results only after this time has elapsed.

The signal ΔU also depends on the velocity of the liquid flow. The results of this experiment are shown in Fig. 6. Both the $\Delta U_L$ and $\Delta U_R$ results show an increase in the voltage value for increasing the liquid flow velocity. The fastest growth rate occurs for the lowest flow rates. For higher values of the flow, a smaller increase in the generated voltage can be seen. On the basis of many experiments, we will assume that this relationship is as in the case of U logarithmic. The values of $\Delta U_L$ and $\Delta U_R$ may be fit with the same logarithmic function[11]:

$$U = a \ln(bv + 1) \qquad (2)$$

where $v$ is flow rate, and constant $a$ and $b$ are: $a = \pm 1.3 \times 10^{-4}$ V and $b = 0.2$ s/m.

In the literature, the topic of dependence of the generated signal on the flow direction in carbon structures (graphene, SWNT) in fluid flow is rarely discussed. There are studies that claim that such an effect does not occur at all [12] or that it does occur, but results are not presented [13]. A flow-dependent result was reported for carbon nanotubes [11]. In this work, the velocity of liquid flow, similarly to our work, was found to fit a logarithmic dependence (2). The logarithmic behavior of the signal describes a typical behavior for motion, which at high speeds includes friction and resistance that lead to a slower increase of the signal for faster flows. The authors [11] explain this phenomenon based on of the Coulomb interaction between the charges in the SWNT layer and in the liquid. Our interpretation of the phenomenon behind the voltage generation is also based on Coulomb interaction, namely by the drag effect. On the other hand the authors of a recent study [38] also observed a logarithmic dependence on the liquid flow velocity and explain this by a flow-induced change of the Fermi level, calculated on the basis of the shifts of the G and 2D bands in the Raman spectra. Both explanations of the logarithmic dependence of the signal can be applied in our system.

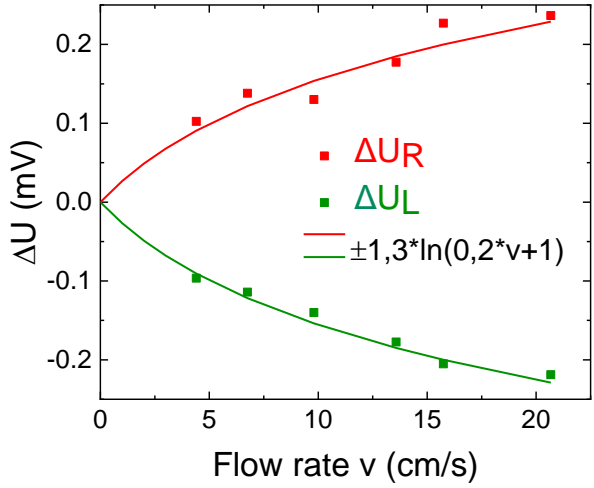

Fig. 6 Dependence of the ΔU_L and ΔU_R signals on the liquid flow rate after applying the procedure, shown in Fig 5. A logarithmic function was fitted to the data.

In the case of graphene exposed to a flowing liquid, we assume that only the following effects may be responsible for signal generation: phonon coupling, Coulomb interactions and streaming potential. Phonon coupling is also known as phonon wind. In this model, the generation of the voltage is by transfer of momentum from the flowing liquid molecules to the acoustic phonons in the nanotube material and dragging free charge carriers in the nanotube material by phonons (phonons wind) [8]. In our case we can reject this mechanism because in such case the flow velocity dependence should be linear, which is not consistent with our results. The phonon coupling theory assumes a linear dependence of the generated voltage on the velocity of the fluid flow, which is clearly inconsistent with our results [8] [9]. Another possible interpretation is connected to the streaming potential. The streaming potential $V_{str}$ is an explanation that applies to micro systems. The relations describing the $V_{str}$ [39] and pressure drop ΔP [40] are:

$$V_{str} = \frac{\varepsilon \varepsilon_0 \zeta \Delta P}{\eta \lambda} \quad (3)$$

$$\Delta P = \frac{8QL\mu}{\pi r^4} \quad (4)$$

where: $\varepsilon$ is the dielectric constant, $\varepsilon_0$ is the vacuum dielectric constant, $\zeta$ is the Zeta potential, $\eta$ is the viscosity, $\lambda$ is the electric conductivity of solution, $Q$ is the flow rate, $L$ is the vessel length, $\mu$ is the viscosity of the fluid, $r$ is the vessel radius. The relations show that ΔP is proportional to the reciprocal of r, which for our macroscopic system without any channel and large distances between electrodes lead to values of ΔP close to zero. The steaming potential $V_{str}$ is proportional to ΔP and therefore we can also claim that this mechanism should not play a significant role in our system.

In our work, which presents signal generation on a macroscopic scale, we propose an explanation involving Coulomb interaction, the so-called "drag effect". A movement of charged particles in solution, via Coulomb forces, affect the charges inside the graphene, dragging them in the direction of flow. Consequently, this effect is dependent on the direction and velocity of the flowing liquid.

The use of additional electrodes allows for better control of the system and the creation of appropriate conditions for the generation of a signal that depends on liquid flow. We show that it is possible to modify these conditions by an external bias voltage. The subtraction of two signals, from the sample and from the solution, allows us to obtain a dependence on both the direction and velocity of the liquid flow. The observed direction-dependent flow signal indicates the importance of the drag effect. The presented results hold great promise for the application of this type of flow sensors, for example, to detect small leakage flows, as well as in microfluidics applications for which a high sensitivity to low flow rates and additional direction sensitivity are of crucial importance.

This work was partially supported by National Centre for Research and Development project GRAF-TECH/NCBR/02/19/2012.